# Molecular Dynamics Simulation of Condensation on Nanostructured Surface in a Confined Space

Li Li[a], Pengfei Ji[b], Yuwen Zhang[b*]

[a] *MOE Key Laboratory of Condition Monitoring and Control for Power Plant Equipment, North China Electric Power University, Beijing 102206, China*

[b] *Department of Mechanical and Aerospace Engineering, University of Missouri, Columbia, MO 65211, USA*

## Abstract

Understanding heat transfer characteristics of phase change and enhancing thermal energy transport in nanoscale are of great interest in both theoretical and practical applications. In the present study, we investigated the nanoscale vaporization and condensation by using molecular dynamics simulation. A cuboid system is modeled by placing hot and cold walls in the bottom and top ends and filling with working fluid between the two walls. By setting two different high temperatures for the hot wall, we showed the normal and explosive vaporizations and their impacts on thermal transport. For the cold wall, the cuboid nanostructures with fixed height, varied length, width ranging from 4 to 20 layers, and an interval of 4 layers are constructed to study the effects of condensation induced by different nanostructures. For vaporization, the results showed that higher temperature of the hot wall led to faster transport of the working fluid as a cluster moving from the hot wall to the cold wall. However, excessive temperature of the hot wall causes explosive boiling, which seems not good for the transport of heat due to the less phase change of working fluid. For condensation, the results indicate that nanostructure facilitates condensation, which could be affected not only by the increased surface area but also by the distances between surfaces of the nanostructures and the cold end. There is an optimal nanosctructure scheme which maximizes the phase change rate of the entire system.

**Nomenclature**

| | |
|---|---|
| $H$ | Height of the modeled system, Å |
| $L$ | Length of the modeled system, Å |
| $r$ | Distance between two atoms, Å |
| $S$ | Surface area |
| $R$ | Surface area ratio |
| $W$ | Width of the modeled system, Å |

*Greek*

| | |
|---|---|
| $\alpha$ | Lattice constant, Å |
| $\sigma$ | Finite distance at which the inter-atomic potential is zero, Å |
| $\varepsilon$ | Depth of the L-J potential, eV |
| $\emptyset$ | Potential function, eV |

*Subscripts and Superscripts*

| | |
|---|---|
| $Ar$ | Argon |
| $Cu$ | Copper |
| $n$ | Cases of simulation |

# 1. Introduction

With the rapid advancement of the society and development of the science and technology, an era with the ubiquitous application of nanotechnologies comes into our daily lives. The integrated circuit inside portable computer, smart phone, and wearable devices has been miniaturized into the nanoscale. The demand of fast processing speed and heavy load of computation lead to higher power consumption from the electronic processing units. Therefore, to efficiently transfer the excessive heat and ensure the normal operation of the nanoelectronic and microelectronic devices have become increasingly urgent [1–3]. Aiming at optimizing the thermal management of height sensitive devices, such as ultrabooks and surface mount circuit board core, flat heat pipe with the thickness less than millimeter provides a promising solution [4–8]. However, the phase change occurring inside the heat pipe still needs to be further understood from a microscopic perspective. Understanding the mechanism of nanoscale and microscale phase change is crucially important to enhance the heat transfer for better thermal management.

The phase change of vaporization and condensation occurring in nanoscale have been hot topics in both theoretical and experimental fields. As a rapidly developing approach to gain sight into the details of intermolecular interaction when phase change happens in nanoscale, molecular dynamics (MD) has been widely recognized and implemented. The thermodynamic information and heat transfer characteristics are obtained by using the MD method in solving the Newton's equation of motion of atom/molecule for a system of interest. Dating back to 1950s, the first computer simulation of MD was started to study systems in both equilibrium and nonequilibrium states [9]. When it comes to study the phase change by using the MD method, Zhang *et al.* [10] carried out MD simulation to understand the evaporation behavior of Lennard-Jones nanodroplets on heated hydrophobic and hydrophilic substrates. The explosive boiling of argon on nanotextured surface was performed by placing different sizes of nanocones, nanoballs on flat substrates [11–13]. The rapid boiling of water on a copper plate was studied by Mao and Zhang [14]. The Kapitza resistance at a water-gold interface in boiling at nanopatterns of varying width-to-spacing ratio and height was studied by Hu and Sun [15]. Both vaporization and condensation of argon on a single platinum surface were simulated by Pan *et al.* [16]. Subsequently, equilibrium molecular dynamics (EMD) and non-equilibrium molecular dynamics (NEMD) were employed by Yu and Wang [17] to study the evaporation and condensation of thin liquid films on a plain wall. A nanoscale evaporating meniscus was simulated to study the disjointing pressure and meniscus boundary condition by Maroo and Chung [18]. Thermal resistance over solid-liquid-vapor interface in a nano-triangular channel was studied by Wang *et al.* [19]. As for investigating the evaporation and condensation rates of liquid at different nanostructured surfaces, Lennard-Jones fluids were modeled as ultra-thin film confined in a nano-channel constituted of two solid walls by Nagayama *et al.* [20].

Most of the aforementioned literature focused on studying the effect of nanostructure to the vaporization and the phenomenon from solid phase to vapor phase in nanosecond order by using the MD simulation. Scant attention was paid to the effect of nanostructure to the phase change from vapor to liquid within the ultrashort time. The condensation on flat surface was seen in [16, 17, 19]. Although the simulation carried out by Nagayama *et al.* [20] considered the condensation on nanostructured surface, it put the emphasis on the phase change on hot wall. Therefore, the present study concentrates on the nanoscale phase change and heat transfer, especially the condensation on the cold wall and the factors affecting the condensation. The classical MD simulation method is adopted. Argon is chosen as the working fluid and copper is used as the material of solid wall. Different temperatures of the hot wall are set up to induce the normal and explosive vaporization on the flat surface. Condensation of the heated argon atoms on cold walls with/without nanostructures are studied. The results will provide a better understanding of the molecular level phase change process.

## 2. Simulation method

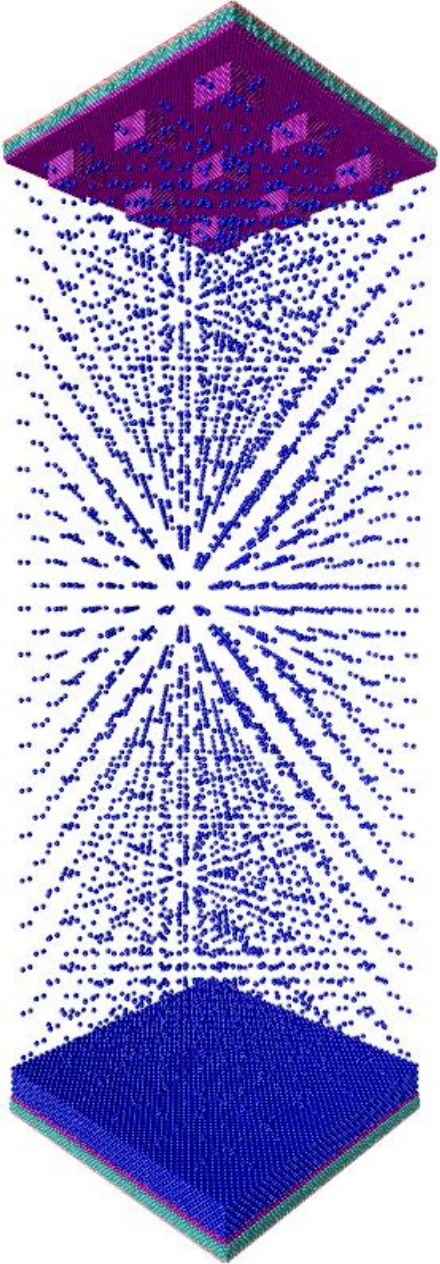 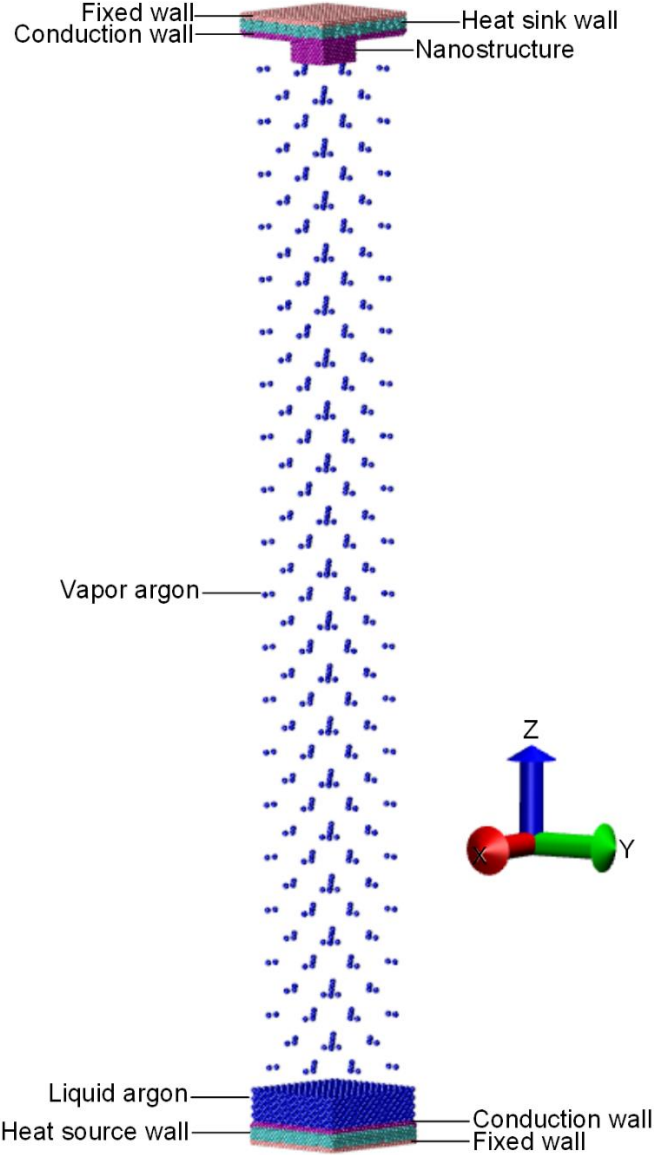

(a) Perspective of nine periodical units      (b) Orthographic view of simulation domain

Fig. 1. Schematic diagram

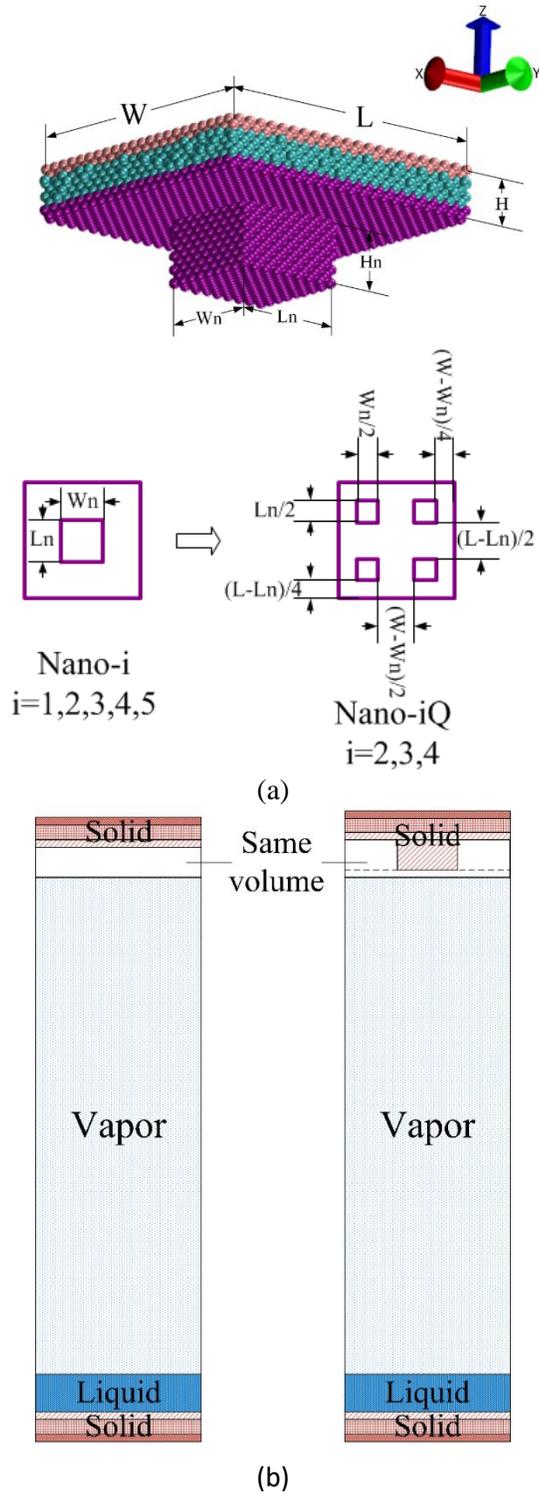

Fig. 2. Schematic diagram of (a) Dimension of nanostructure and (b) Spatial arrangement of simulation domain

The molecular system under consideration is consisted of solid copper walls at the two ends and argon with liquid and vapor phases filled between the two copper walls (see Fig. 1). The simulation was performed by using the Large-scale Atomic/Molecular Massively Parallel Simulator (LAMMPS) under the framework of classical MD [21]. The VMD software [22] was used to obtain the visual graphs of system configuration and atomic motion. Lattice structure of face-centered cubic (FCC) unit cells were used to model the solid walls and the liquid/vapor working fluid. Periodic boundary condition was applied in the $x$- and $y$- directions. The total length (in the $x$-direction) and width (in the $y$-direction) of the simulation box were set as 22 layers. The height (in the $z$-direction) of the box was the summation of the heights of all the materials.

For the copper wall, the lattice constant of FCC unit cell was set as 3.6149 Å, which enables the conversion of density of copper as $8.96 \, g/cm^3$. Solid copper walls with height of 4 layers were placed on both bottom and top of the simulation box. For both the hot and cold walls, they were divided into fixed wall, heat source/sink wall and heat conduction wall. Copper atoms in the fixed wall were fixed without atomic vibration, which prevent the thermal energy of the entire system from losing and atoms from flying away. The heat source/sink wall were adjacent to the fixed wall, which were controlled by the Berendsen thermostats [23] for heat flowing from the hot side to the cold side. The control of temperature was realized by rescaling velocities of copper atoms locating in the heat source/sink. The heat conduction walls were implemented to conduct heat from the heat source to the liquid argon and absorb thermal energy from the condensed argon to the heat sink. From the perspective of heat transfer, constant high temperature boundary condition was placed on the hot wall to supply the heat to the liquid argon for the vaporization, while constant low temperature boundary condition was imposed on the cold wall for the condensing the heated argon from the hot wall. Two kinds of temperature treatments with $130 \, K$ (hot wall) versus $85 \, K$ (cold wall) and $300 \, K$ (hot wall) versus $85 \, K$ (cold wall) were set up. For each treatment, the hot wall was modeled as flat surface, whereas the cold wall was configured with cuboid nanostructures located on the center of the flat surface.

The nanostructures were constructed with the same height of 5 layers and five different numbers of layers in $x$- and $y$- directions ($L_n \times W_n = 14.46 \times 14.46, \; 28.92 \times 28.92, \; 43.38 \times 43.38, \; 57.84 \times 57.84, \; 72.3 \times 72.3$ Å). For convenience, the above five nanostructures were denoted as Nano-1, Nano-2, Nano-3, Nano-4 and Nano-5. In addition, three different nanostructures arranged as 4 quarters of Nano-2, Nano-3, Nano-4 with the same height of 5 layers were designed to further illustrate the effect of the nano-shape and they were denoted as Nano-2Q, Nano-3Q, and Nano-4Q (see Table 1 and Fig. 2(a)). The ratio of the entire surface area of condensation in Nano cases to the surface area in flat case is defined as $R_0$.

At the initial stage of simulation, the liquid argon region were in contact with the hot wall side, which includes 3,136 atoms with the arrangement of FCC unit cells. The lattice constant of liquid argon was chosen as 5.7441 Å, which agreed with the density of liquid argon $1.40 \, g/cm^3$ at its boiling point of $87.3 \, K$ at $1 \, atm$. Above the region of the liquid argon, 487 vapor argon atoms with the lattice constant equal to 36.2507 Å for the FCC unit cells were placed, which ensure the density of the modeled vapor argon equal to the $5.57 \, g/L$ at its boiling point $87.3 \, K$ at $1 \, atm$. Because of the mismatch between the lattice constant of vapor argon and copper, there will be discrepancies among the total numbers of vapor argon atoms if they are occupied into the space between liquid argon and cold conduction wall mounted with different sizes of nanostructures. During the modeling process, an empty space was left to ensure the number of vapor argon atoms are the same for all the cases (see Fig. 2 (b)). In other words, once the nanostructure changes, its occupied space is compensated by increasing the height of the aforementioned space.

Table 1. Structures of solid atoms at condenser site

| Nanostructure configurations | z-view | $S$-Total surface area at condenser site (Å × Å) | $R_0$-Surface ratio $\dfrac{S}{22^2 \times 3.6149^2}$ |
|---|---|---|---|
| Flat surface | 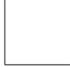 | $22^2 \times 3.6149^2 = 6324.67$ | 1 |
| Nano-1 | 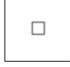 | $(4 \times 4 \times 5 + 22^2) \times 3.6149^2 = 7370.07$ | 1.165 |
| Nano-2 | 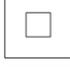 | $(4 \times 8 \times 5 + 22^2) \times 3.6149^2 = 8415.47$ | 1.331 |
| Nano-3 | 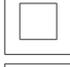 | $(4 \times 12 \times 5 + 22^2) \times 3.6149^2 = 9460.87$ | 1.496 |
| Nano-4 | 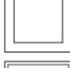 | $(4 \times 16 \times 5 + 22^2) \times 3.6149^2 = 10506.27$ | 1.661 |
| Nano-5 | 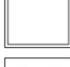 | $(4 \times 20 \times 5 + 22^2) \times 3.6149^2 = 11551.67$ | 1.826 |
| Nano-2Q | 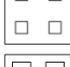 | $(4 \times 4 \times \left(\dfrac{8}{2}\right) \times 5 + 22^2) \times 3.6149^2 = 10506.27$ | 1.661 |
| Nano-3Q | 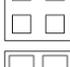 | $(4 \times 4 \times \left(\dfrac{12}{2}\right) \times 5 + 22^2) \times 3.6149^2 = 12597.07$ | 1.992 |
| Nano-4Q | 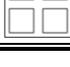 | $(4 \times 4 \times \left(\dfrac{16}{2}\right) \times 5 + 22^2) \times 3.6149^2 = 14687.87$ | 2.322 |

Because the Lennard-Jones (LJ) fluid has a good representation of phase transitions between vapor-liquid, solid-liquid and solid-vapor phases, as well as the critical points [24, 25], it was chosen to model and simulate the properties of real fluids. The interatomic interaction between copper-copper, argon-argon and argon-copper atoms were described by the standard $12 - 6$ LJ potential function with a cutoff distance of $3.7\ \sigma_{Ar}$, which had already shown to be a simple and reliable potential function. Parameters used for atomic interactions are listed in Table 2 [26]. The parameters of $\varepsilon_{Cu-Ar}$ and $\sigma_{Cu-Ar}$ were calculated according to the Lorentz-Berthelot mixing rule [27], namely, $\varepsilon_{Cu-Ar} = \sqrt{\varepsilon_{Cu}\varepsilon_{Ar}}$ and $\sigma_{Cu-Ar} = (\sigma_{Cu} + \sigma_{Ar})/2$.

Table 2. L-J potential parameters for argon-argon, metal-metal and argon-metal atoms [25]

| $\emptyset = 4\varepsilon[(\dfrac{\sigma}{r_{i,j}})^{12} - (\dfrac{\sigma}{r_{i,j}})^{6}]$ | Argon-Argon | Copper-Copper | Argon-Copper |
|---|---|---|---|
| $\varepsilon$ (eV) | 0.0104 | 0.4096 | 0.0653 |
| $\sigma$ (Å) | 3.4050 | 2.3400 | 2.8725 |

The simulation process can be divided into three consequent stages. For the entire simulation, the time step was set as $1\ fs$. The first stage was to perform energy minimization of the entire system. The minimization process was performed by using conjugate gradient (CG) algorithm. The stopping tolerance for energy and force were taken as $1.0 \times 10^{-5}\ eV$ and $1.0 \times 10^{-6}\ eV/\text{Å}$, respectively. The number of steps was limited in 10,000 steps. At the second stage, uniform temperature of the liquid and vapor was set as $87.3\ K$, respectively. The canonical ensemble was used to keeping number of atoms, constant volume and constant temperature (NVT). Temperature of the heat source wall was set as $130\ K$ for the normal vaporization and $300\ K$ for the explosive vaporization. Temperature of the heat sink wall was set as $85\ K$. The temperature controls were realized by implementing the Berendsen thermostats [28] for $0.1\ ns$. Subsequently, the microcanonical ensemble simulation was carried out to argon by keeping number of atoms, constant volume and constant energy (NVE) for another $0.1\ ns$. Meanwhile, the walls of heat source and heat sink were operated under constant temperatures. During the second stage, the heat conduction walls were set as adiabatic wall to impede heat exchange between the heat source/sink and working fluid. Temperature evolutions of the wall (copper atoms) and working fluid (argon atoms) show stable profiles around $85\ K$ and $87.3\ K$, which indicated the liquid/vapor argon has reached sufficiently thermal equilibrium. At the third stage, the adiabatic control of the heat conduction wall contacting the heat source/sink walls and working fluid was removed, which allowed transfer the heat from hot wall to cold wall through phase change of the working fluid. Nose-Hoover thermostats [29,30] were added to maintain the temperatures of the heat source/sink, which are the same as those at the second stage. Compared to Berendsen thermostat in rescaling the velocities of atom to control the temperature, Nose-Hoover thermostat is able to correctly generate trajectories consistent with a canonical ensemble. Throughout the above three stages of simulations, the temperatures of walls, numbers and trajectories of liquid/vapor atoms were recorded to analyze the transport phenomenon.

## 3. Results and discussion

### 3.1 Effect of nanostructure size and arrangement

Figure 3 illustrates the average temperature of the argon atoms, including liquid and vapor phases. Immediately after removing the adiabatic control of the heat conduction wall, the average temperatures instantly respond to values slightly below $130\ K$ (for the heat conduction wall near the heat source wall) and greater than $85\ K$ (for the conduction wall near the heat sink wall). Due to the temperature difference between the copper atoms of the conduction wall and the argon atoms contacting with the conduction wall, the initially placed liquid argon atoms are heated up. The instant heating of the liquid argon leads to sharp increase of its temperature to nearly $130\ K$. By further heating the liquid argon, argon atoms in the upper position of the liquid region overcome the attractive force interacting with the liquid argon and start to vaporize.

The argon atoms in vapor state sourcing from the liquid argon continue to move towards the cold wall. During the course of the moving, the atomic collision between the newly generated vapor argon atoms and the initially existing vapor argon atoms occurs, which results in the transport of thermal energy between the vapor argon atoms. The thermal energy transport continuously takes place until the vapor atoms with high energy arrive at the conduction wall in the cold side and release the remained energy to cold wall. The vapor argon atoms begin to condense at the surface of the cold conduction wall. The argon atoms continuously accumulate near the conduction wall of the cold side, which reduces the number of argon atoms being heated. Therefore, the average temperature of argon begins to slowly decrease. Since the condensation of vapor argon takes place after it passes a relatively long distance, the decrease process is slower than the steep increase from since $0.2\ ns$. During the transient phase change process, liquid argon at the hot side absorbs heat and vaporizes as a result of temperature increase. Meanwhile, the vapor argon at the cold side releases heat and condenses as a result of temperature decrease. When the average of temperature of all argon atoms decreases, it demonstrates that the condensation plays the dominating role. The introduction of nanostructure expands the area of the cooling, which helps to enhance the local heat transfer near the cold wall for some extent. As seen in the inserts ($0.8\ ns - 1.0\ ns$ and $2.8\ ns - 3.0\ ns$) in Fig. 3, the temperatures of those including nanostructures show lower values than that of the flat surface. In Fig. 3, the six temperature evolutions present resemble curves for the cases of nanostructures and flat surface. At the end of the simulation, the temperature differences among the cases with and without nanostructures get smaller than those during $0.8\ ns - 1.0\ ns$.

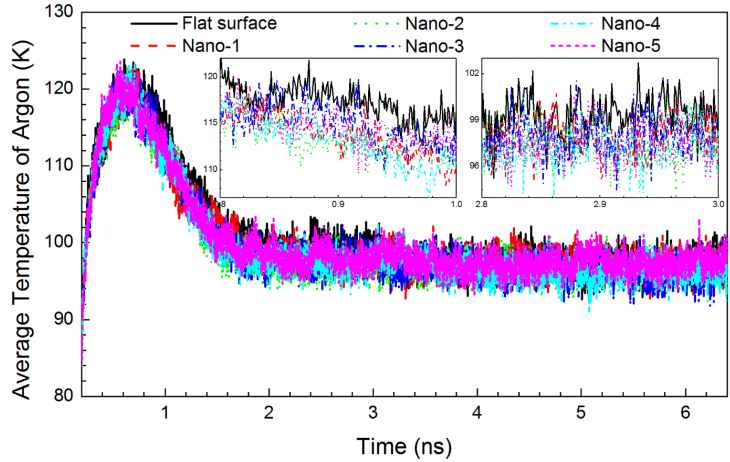

Fig. 3. Temperatures of argon atoms for Nano-i structures for hot wall ($130\ K$) versus cold wall ($85\ K$)

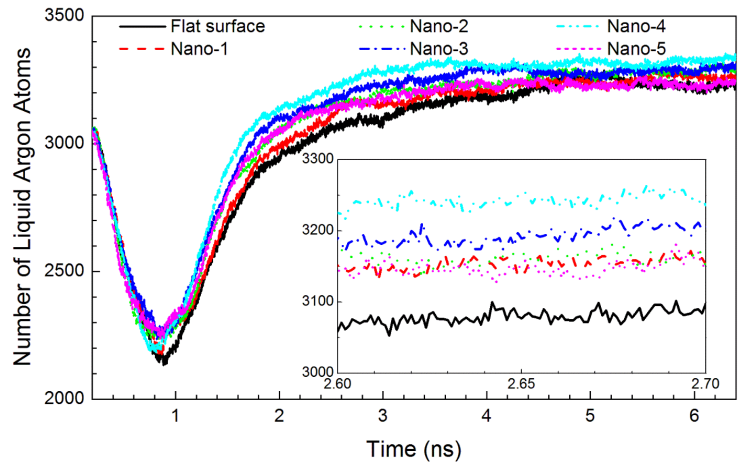

Fig. 4. Numbers of liquid argon atoms for Nano-i structures for hot wall ($130\ K$) versus cold wall ($85\ K$).

Figure 4 shows the variations of the number of liquid argon atoms for different cases. At the beginning, the number of liquid atoms rapidly decrease because of the instantly heated liquid atoms start to leave the liquid region and move into vapor region. Although on the cold wall side the higher temperature vapor atoms transfer energy to lower temperature copper atoms and condensation occurs, the number of atoms of vapor phase changing from liquid phase is much larger than that of liquid atoms changing from vapor phase in this stage. After $0.8\ ns$, the number of liquid argon atoms starts to increase due to the condensation of more and more vapor. After $2.0\ ns$, the numbers of liquid argon atoms in all cases gently increase because the stable layer of condensed argon are formed on the surface of the cold conduction wall. Compared with the cases with nanostructure, the flat surface case shows the lowest profile because of the smallest surface area of argon contacting with the cold wall. The noticeable difference of transport phenomena between nanostructured surface emerge gradually. For the six cases of hot wall ($130\ K$) versus cold wall ($85\ K$), the number of liquid argon atoms increases as the surface area ratio (which is defined as the ratio of surface area with nanostructure to the surface area of flat wall) increases up to 1.661 (Nano-4 case), then the number decreases when the ratio of surface area is 1.826 (in Nano-5 case). Although the ratio of Nano-5 case is larger than that of Nano-4 case, the space around the nanostructure is too small to allow large amount of argon atoms to be condensed, which makes the total effect closer to a flat surface case.

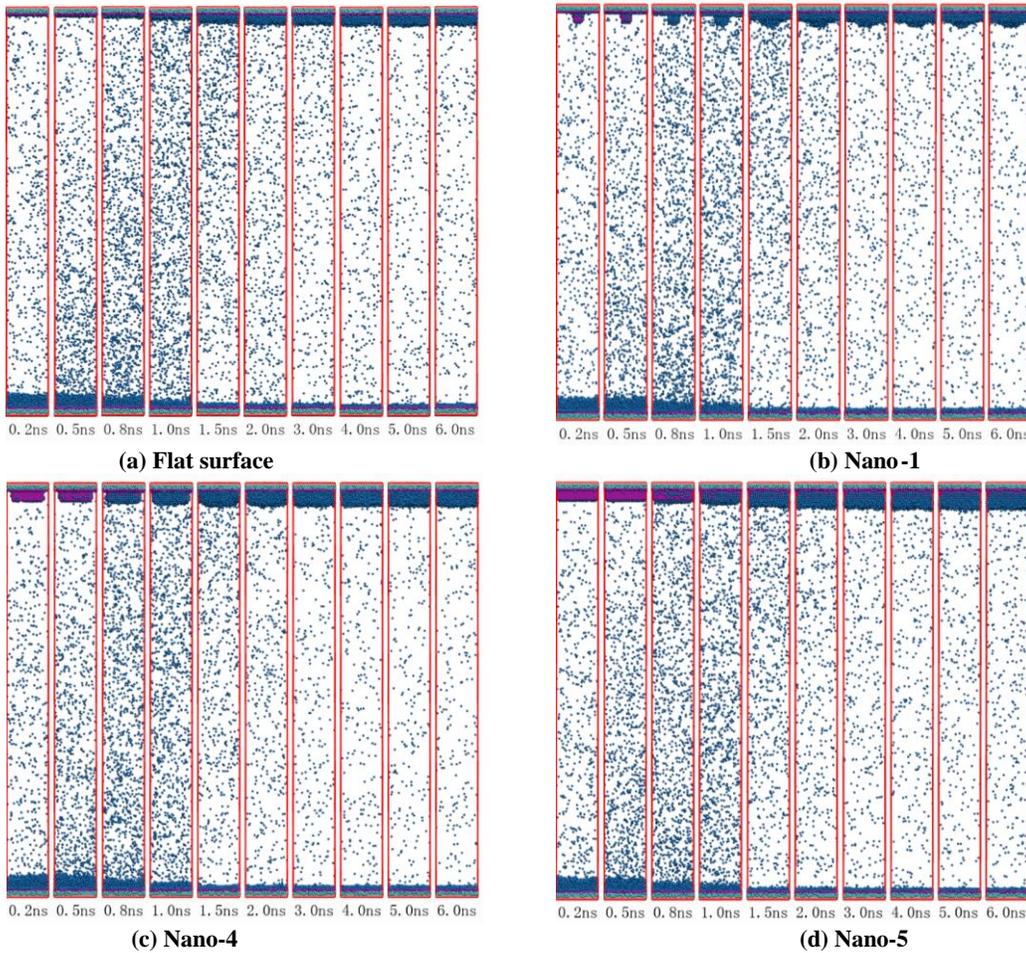

Fig. 5. Trajectories of atoms for nanostructures (x view) for hot wall ($130\ K$) versus cold wall ($85\ K$). (a) Flat surface, (b) Nano-1, (c) Nano-4, (d) Nano-5

Figure 5 shows the *y-z* projections of simulation domains of four cases (Flat surface, Nano-1, Nano-4 and Nano-5) at different times. At the bottom part of the working fluid region, the interface between liquid argon and vapor argon is clearly visible. As described in Section 2, since $0.2\ ns$, heat conduction begins from the heat source wall to liquid argon through the conduction wall. With the heating of the liquid argon atoms, the vaporization starts from the surface of liquid region. Some of the argon atoms convey thermal energy by leaving from the liquid region to the vapor region. On the other hand, at the top part a few vapor argon atoms begin to condense on the surface of cold conduction wall. At the beginning of heat flux flowing from the hot end to the cold end, the number of liquid argon atoms decreases, meanwhile that of vapor argon atoms increases. The enlarged surface area by nanostructure absorbs more thermal energy, which causes the condensation of more vapor atoms.

In Figs. 5 (a) -(c) at $0.5\ ns$, it is interesting to notice that the condensation firstly occurs on the base flat surface and the surrounding surfaces of nanostructure in the cases of flat surface, Nano-1 and Nano-4. While for the case of Nano-5, the condensation primarily occurs on the external flat surface (parallel to the $x-y$ plane) of nanostructure, as shown in Fig. 5 (d). The expanded surface area plays a major role in the effect at the beginning of phase change. Comparing Fig. 5 (a) and (b) from $1.0\ ns$ to $2.0\ ns$, it can be seen that the number of liquid argon atoms aggregating on the cold wall with nanostructure is greater than that without nanostructure. Comparing Figs. 5 (a)-(c) at time of $2.0\ ns$, for the cases of Nano-1 and Nano-4, the numbers of liquid argon atoms condensed on cold wall are more than that on flat surface case, which can be explained by the greater rate of condensation caused by the increased area of cooling. Nano-2 and Nano-3 cases have almost the same features as Nano-1 and Nano-4, so that the snapshots for these two cases are not shown. Whereas, for the case of Nano-5, the number of liquid atoms on the surface of cold wall is not more than that of Nano-4. As shown in Fig. 5 (d), there are still some regions of cold end cannot be covered by the liquid argon atoms. The special result of condensed liquid argon atoms for the case of Nano-5 gives a visual explanation that why there is greater number of liquid argon in the case of Nano-4, even though there is large cooling area for Nano-5 than that for Nano-4.

With the gradually decreasing number of vapor argon atoms between the hot and cold ends, the total number of liquid argon atoms increases slowly. Additionally, some of the argon atoms stay at the bottom as a thin film and get vaporized slowly in the subsequent snapshots. The similar result that an ultra-thin layer of argon atoms staying on the hot side no matter how long the process last was reported in [31]. They concluded that the thickness of the non-evaporating liquid film decrease with the increase of heating substrate temperature, which is consistent with our results that will be shown in section 3.2 for the cases with explosive boiling. Physically, the fluid atoms staying on the surface of wall is caused by the strong intermolecular forces [17].

In order to get further understanding of the phase change between vapor and liquid quantitatively, it is important to investigate the number of spatial distribution of argon atoms. Figure 6 shows the number of argon atoms distributed along the $z$-direction at different time. Large numbers of argon atoms crowd near the two ends, which demonstrates the existence of liquid argon. In each subfigure of Fig. 6, the evolution of vaporization and condensation near the hot wall and cold wall could be clearly seen by comparing the curves at different times. At beginning of condensation on the cold surface, the number density of argon atoms near a cold wall gradually appears and increases. Meanwhile, the number density of liquid atoms layers near the hot wall decreases. For Fig. 6 (a), the number density curves show the peaks at the interface with the cold wall. Whereas, for Figs. 6 (b)-(f), because of the occupation of the nanostructures solid atoms, as the increasing of surface ratio, the peak of number density near cold wall move to lower position in the $z$-direction. The space between peaks near hot and cold ends are full of

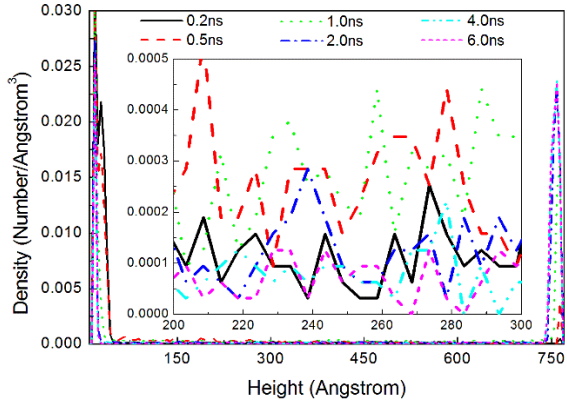
**(a) Flat surface**

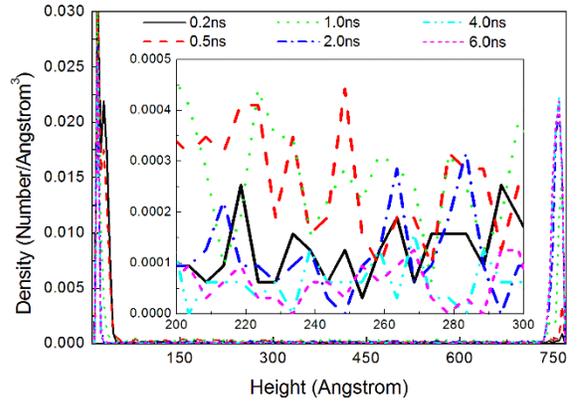
**(b) Nano-1**

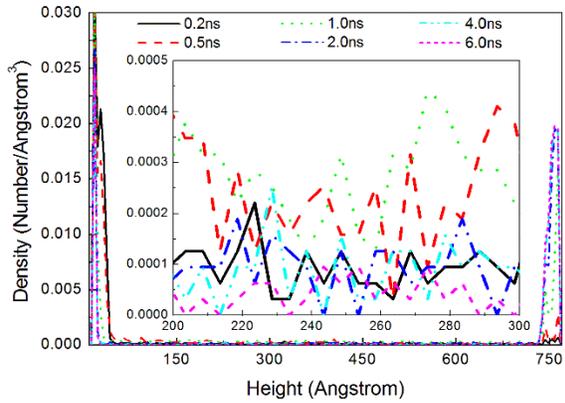
**(c) Nano-2**

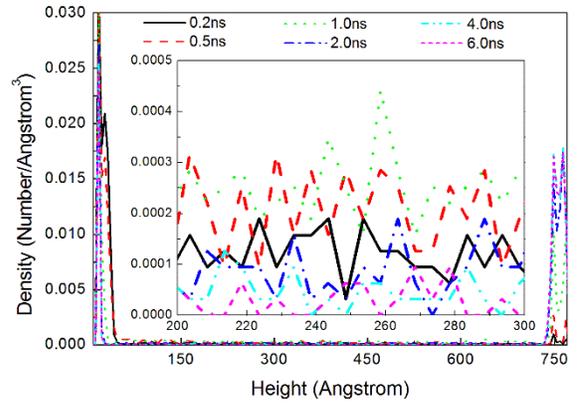
**(d) Nano-3**

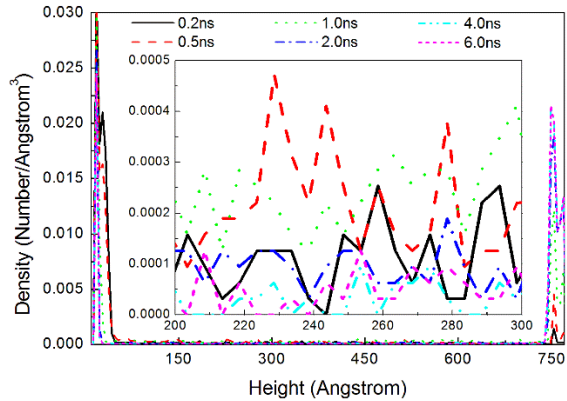
**(e) Nano-4**

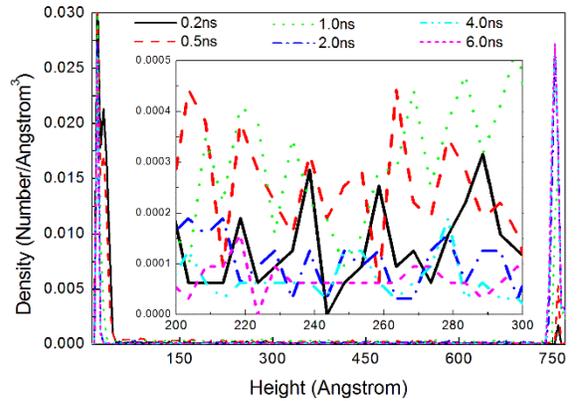
**(f) Nano-5**

Fig. 6. Number densities of argon atoms for Nano-i structures for hot wall ($130\ K$) versus cold wall ($85\ K$). (a) Flat surface, (b) Nano-1, (c) Nano-2, (d) Nano-3, (e) Nano-4, (f) Nano-5

amount of vapor atoms which with low values of number density. From the magnified graphs in Figs. 6 (a)-(f), the number density of vapor atoms rises firstly (from $0.2\ ns$ to $1.0\ ns$) then drops down. It means during the whole process of simulation, the vaporization play the dominant role firstly. Subsequently, the vapor number being condensed increases to balance the vapor number being generated. Finally, the condensation causes the total number of vapor atom decreases. It can be seen from Fig. 6 (f), due to nano-5 structure has the narrowest distance between adjacent surfaces, it results in a sharp decrease of the number density near base cold wall.

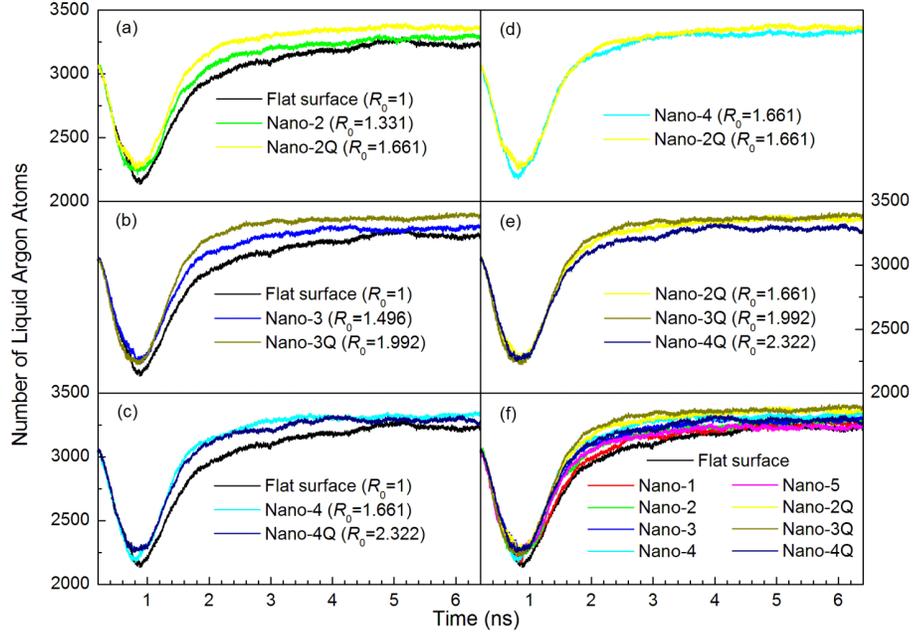

Fig. 7. Numbers of liquid argon atoms for Nano-iQ structures for hot wall (130 $K$) versus cold wall (85 $K$). (a) Nano-2 and Nano-2Q, (b) Nano-3 and Nano-3Q, (c) Nano-4 and Nano-4Q, (d) Nano-4 and Nano-2Q, (e) Nano-iQ, (f) All structure cases

To investigate whether the larger area of nanostructure is the better for phase change and the importance of role of space between adjacent surfaces plays, three different nanostructures with 4 quarters were investigated; the results were shown in Fig. 7. The comparisons among Nano-2, Nano-3, Nano-4 and Nano-2Q, Nano-3Q, Nano-4Q, which have the same nanostructure area of projection in z-direction between Nano-i and Nano-iQ (i is index of 2, 3 and 4) in each subfigures (a) ~ (c), are shown. Comparing between Figs. 7 (a) and (b), the number of liquid argon atoms can be seen nearly 7% in Nano-iQ structures more than that in Nano-i structures. Nano-3Q got the most number of liquid atoms among all of the cases. In contrast, from Fig. 7 (c), the number of liquid argon atoms decreased after the nano-4 structure had been divided into 4 pieces in nano-4Q case which has the largest surface area but smallest space between adjacent surfaces. Figure 7(d) presents the two cases with the same surface area ratio. It is seen that Nano-2Q showed slightly greater values than the Nano-4 case. In Fig. 7(e), Nano-2Q and Nano-3Q can get the larger number of liquid atoms during the instant condensation than that of Nano-4Q, which can conclude that the more uniform space (neither too big nor too small) between adjacent surfaces is the dominant factor, rather than the larger surface area.

For the purpose of quantify the impact of adjacent spaces, the ratio of the increased surface area of condensation in Nano cases to the surface area in flat case was defined as $R_1$, the ratio of the total nano surface area of x-z and y-z projections to the cold flat wall subtract nano-structures was defined as $R_2$ and the reciprocal of $R_2$ was defined as $R_3$, as shown in Table 3. It can be seen that the ratio of the adjacent surfaces in the range of nearly 0.7 ~ 1.5 may be the optimal scheme from the values of $R_2$ and $R_3$. In this range, the effect of condensation enhanced with the increasing of $R_1$.

Table 3. Surface area ratio at condenser site

| Nanostructure configurations | $S_{n,xy}$ (see shade part) Nano surface area in x-z & y-z projections 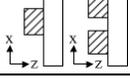 $R_1$-Surface area ratio $(\frac{S_{n,xy}}{S_{n,z}+S_{f-n,z}})$ | $S_{n,z}$ (see shade part) Nano surface area in x-y projection 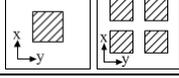 $R_2$-Surface area ratio $(\frac{S_{n,xy}}{S_{f-n,z}})$ | $S_{f-n,z}$ (see shade part) Surface area subtract Nano surface area in x-y projection 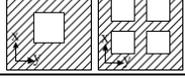 $R_3$-Surface area ratio $(\frac{S_{f-n,z}}{S_{n,xy}})$ |
|---|---|---|---|
| Flat surface | 0 | 0 | 0 |
| Nano-1 | 0.165 | 0.171 | 5.848 |
| Nano-2 | 0.331 | 0.381 | 2.625 |
| Nano-3 | 0.496 | 0.704 | 1.42 |
| Nano-4 | 0.661 | 1.404 | ⓪.712 |
| Nano-5 | 0.826 | 4.762 | 0.21 |
| Nano-2Q | 0.661 | ⓪.762 | 1.31 |
| Nano-3Q | 0.992 | 1.412 | ⓪.708 |
| Nano-4Q | 1.322 | 2.807 | 0.356 |

## 3.2 Effect of temperature of hot wall

In order to research the effect of temperature on evaporation and condensation, the heat source wall was set as $300\ K$, which was far away from the critical temperature of the argon (150.7K). The evolution of average temperature of argon atoms is shown in Fig. 8. The temperature of argon atoms rises rapidly at the beginning. A steeper slope of temperature profile is seen, which demonstrates the response of argon temperature in explosive vaporization is faster than that in the normal vaporization. The peak is induced by instant heating of the liquid argon contacting with the hot conduction wall. The descending profile after the peak (near $0.3\ ns$) is induced by the separation of the argon cluster from the hot wall (which can be seen in the latter Fig. 12). The vapor region between the separated argon part and the heat conduction wall weakens the transfer of the heat, which leads to the faster fall of the average temperature of argon than that in Fig. 3. After the rapid decline from $0.3\ ns$ to $1.0\ ns$, the average temperature declines slowly. After $1.0\ ns$, it is worth to notice that the average temperature of argon atoms is around $100\ K$ although the temperature of hot wall is up to $300\ K$.

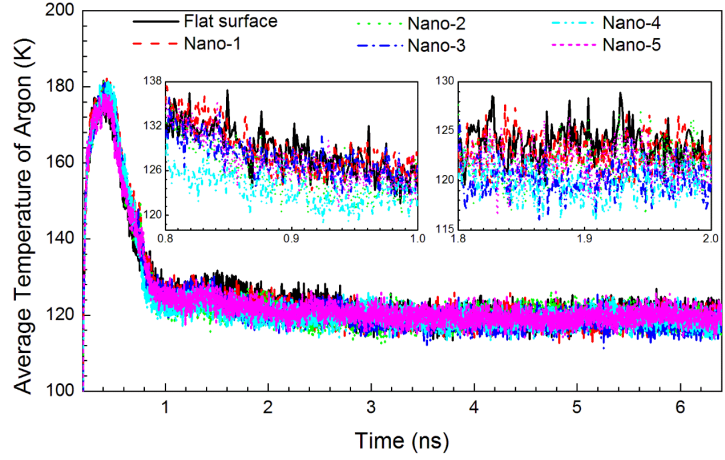

Fig. 8. Temperature of argon atoms for Nano-i structures for hot wall (300 $K$) versus cold wall (85 $K$)

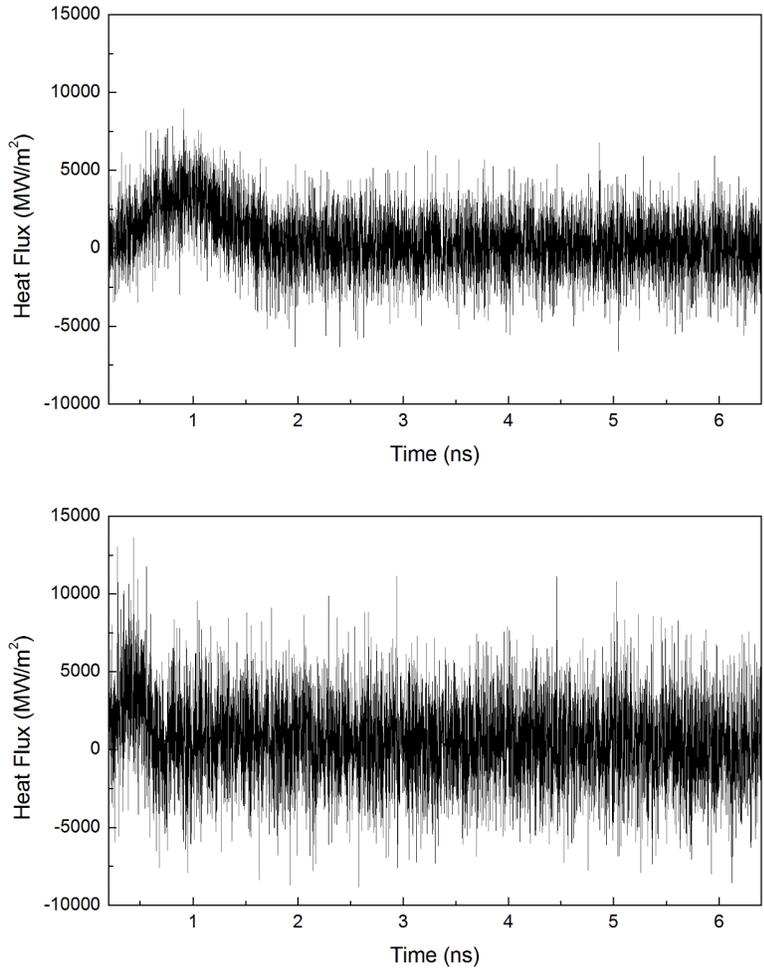

Fig. 9. Heat flux of initial liquid argon region for the flat surface case. (a) At low temperature (130 $K$), (b) At high temperature (300 $K$)

The explosive vaporization leads to the cluster of argon atoms get separated from the hot conduction wall, which is equivalent to reduce the number of working fluid contacting with the hot wall. Figure 9 (a) and 9 (b) are the heat flux graphs of the initial liquid argon region near the bottom end at $130\ K$ and $300\ K$, respectively. The heat flux of the $300\ K$ case gets greater value than that of the $130\ K$ case during the first stage from $0.2\ ns$ to $0.4\ ns$. It drops suddenly to approximate average zero after $0.6\ ns$. However, the heat flux of $130\ K$ increases steadily from $0.2\ ns$ to $1.0\ ns$ and then drops gradually to approximate average zero in the latter $1.0\ ns$. These phenomena can proved that in the $300\ K$ case, the large majority of argon atoms move too fast to have time to convey the heat away from the hot side in the ultra-short time. In order to improve the efficiency of heat transfer, measures should be taken to reduce the occurrence of large scale separation of the working fluid.

Various sorts of nanostructures designed on hot wall to study the vaporization are reported in [11–13]. Nevertheless, as shown in [11–13], the initial height of liquid atoms on the hot end without nanostructure is the same as the height of case with nanostructure. It means the amount of liquid atoms are not the same between the cases with and without nanostructures. That is to say the initial number of liquid atoms are reduced when nanostructure occupied the volume. Therefore, the conclusion that the placement of nanostructures on the hot wall could enhance heat transfer should be further investigated. More discussions need to be carried out under the condition that the same amount of the working fluid in order to investigate the effect.

Although liquid cluster is not good for conveying more amount of heat, for the perspective of cycling of the working fluid, the liquid cluster may reduce the time cost in a cycle. As shown in Fig. 4 and the latter Fig. 10, for the case of $130\ K$ versus $85\ K$, the time cost (for most of the argon get condensed in the cold end) is more than 2 times than that for the case of $300\ K$ versus $85\ K$. Figure 10 shows the number of liquid argon atoms in the Nano-i cases and flat surface case. Figure 11 shows the comparisons between Nano-i structures and the Nano-iQ structures. Comparing with the results of $130\ K$ versus $85\ K$ case, the total time need to reach equilibrium for the case of $300\ K$ versus $85\ K$ is shorter. The same conclusion of the enhanced effect of nanostructures can be drawn but the difference between nanostructure surfaces is not obvious as those in $130\ K$ versus $85\ K$ case.

The snapshots (see Fig. 12) are different from those in Fig. 5. For the explosive vaporization, most of the argon atoms move as a cluster, rather than individually dispersed. Because of the great difference of temperature and heat flux between the wall and the adjacent liquid argon, a small proportion of liquid argon is heated instantly as a result of explosive vaporization. The vaporization firstly occurs in the lower position of the liquid region. The great pressure caused by the expansion of the vaporized atoms pushes the upper atoms (which are in liquid phase) to move up as a cluster. The aggregated argon atoms separate from the vaporized argon atoms and move upward. Additionally, some of the argon atoms stay at the bottom as a thin film and get vaporized slowly in the subsequent snapshots. During the process of the cluster moving from the hot end to the cold end, more and more argon atoms leave from the bottom of the cluster because of the continuous heating and thermal expansion of the argon cluster after separating from the hot wall. Due to the pressure gradient along the $z$- direction, the cluster stays at surface of the cold conduction wall and nanostructures, after arriving at the conduction wall in the cold end. The remained vapor atoms condensate on the bottom surface of the cluster. Owing to the continuous cooling from the cold conduction wall, the condensed liquid argon get tighter and tighter (see snapshots at $1.0\ ns$). The snapshots in the case of Nano-4 show faster speed of upward movement of cluster.

The number densities for the cases with different nanostructures are shown in Fig. 13. From the location of peak at different times, it can be seen that the liquid cluster move from the hot end to the cold end. The height of peaks in Fig. 13 reduce with the progress of simulation. The changes of peaks indicates argon atoms around the cluster leave away from the cluster gradually. The velocity of cluster for the cases with nanostructures is faster. The placement of nanostructure makes the happening of condensation near the cold wall earlier. Nano-4 shows the best amount of condensation near the cold wall than others, which agrees with the conclusion for the case of 130 $K$ versus 85 $K$.

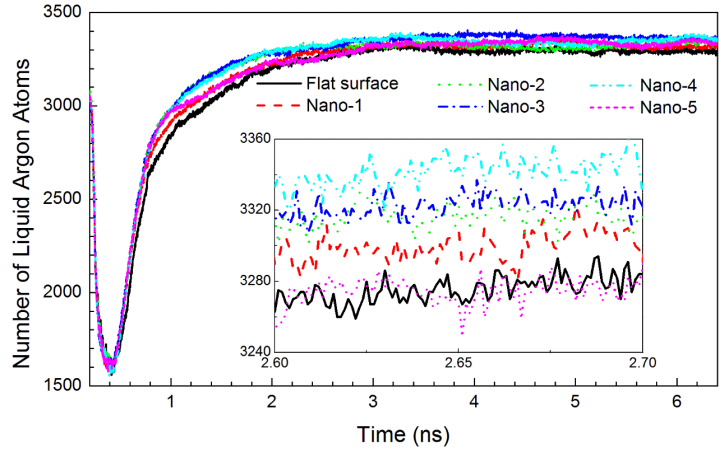

Fig. 10. Numbers of liquid argon atoms for Nano-i structures for hot wall (300 $K$) versus cold wall (85 $K$)

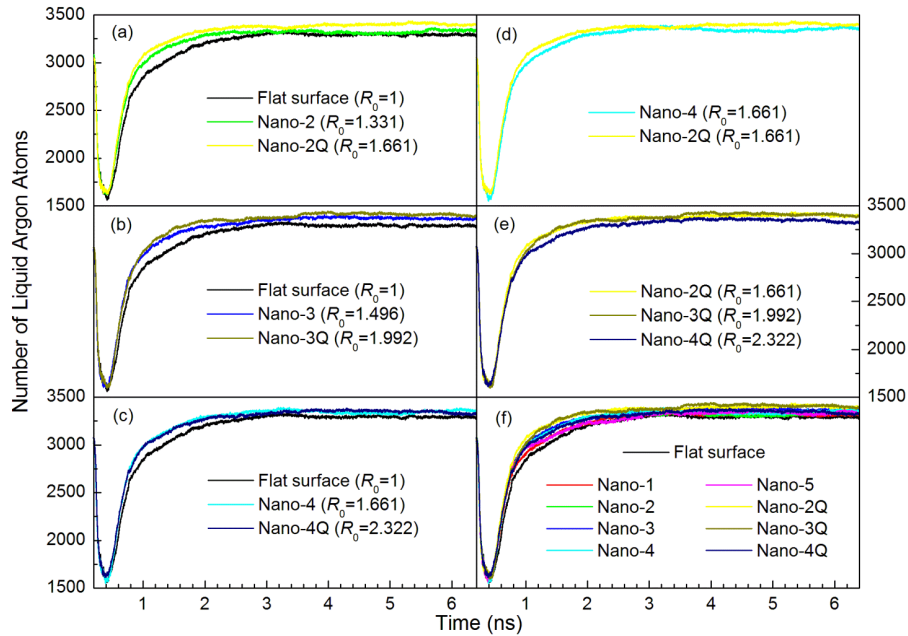

Fig. 11. Numbers of liquid argon atoms for Nano-iQ structures for hot wall (300 $K$) versus cold wall (85 $K$). (a) Nano-2 and Nano-2Q, (b) Nano-3 and Nano-3Q, (c) Nano-4 and Nano-4Q, (d) Nano-4 and Nano-2Q, (e) Nano-iQ, (f) All structure cases

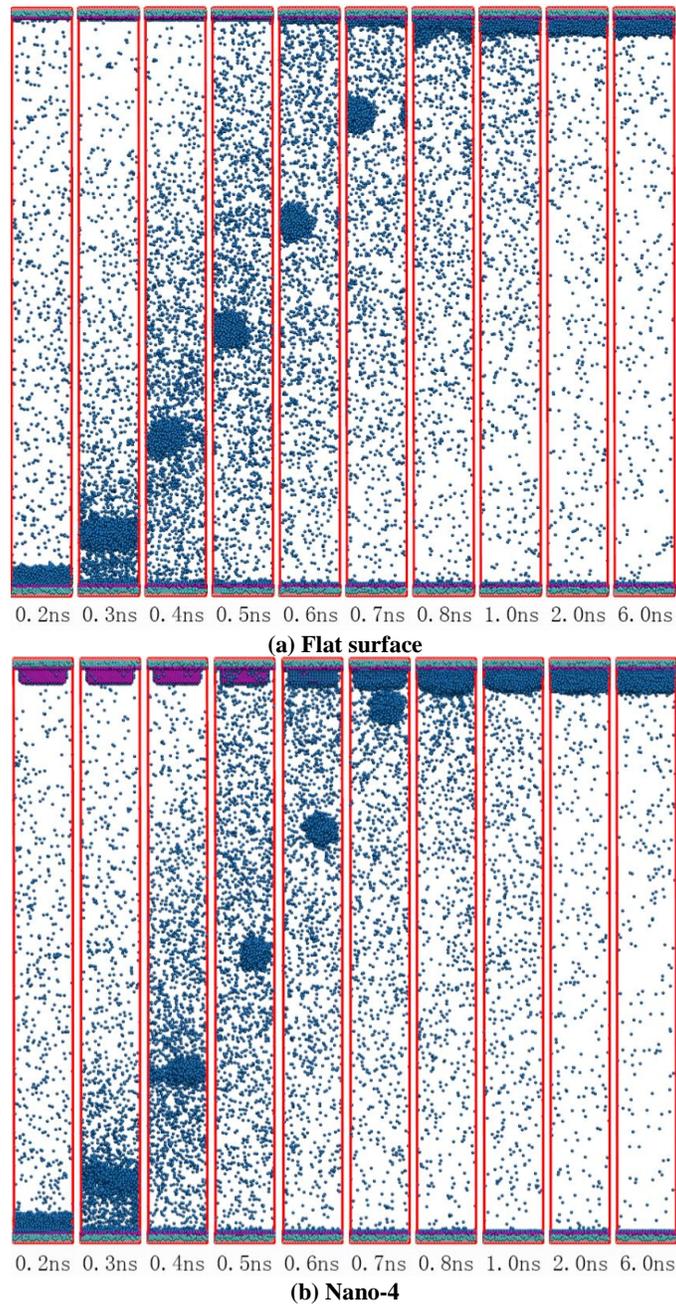

(a) Flat surface, (b) Nano-4

Fig. 12. Trajectories of atoms for nanostructures (x view) for hot wall (300 $K$) versus cold wall (85 $K$).
(a) Flat surface, (b) Nano-4

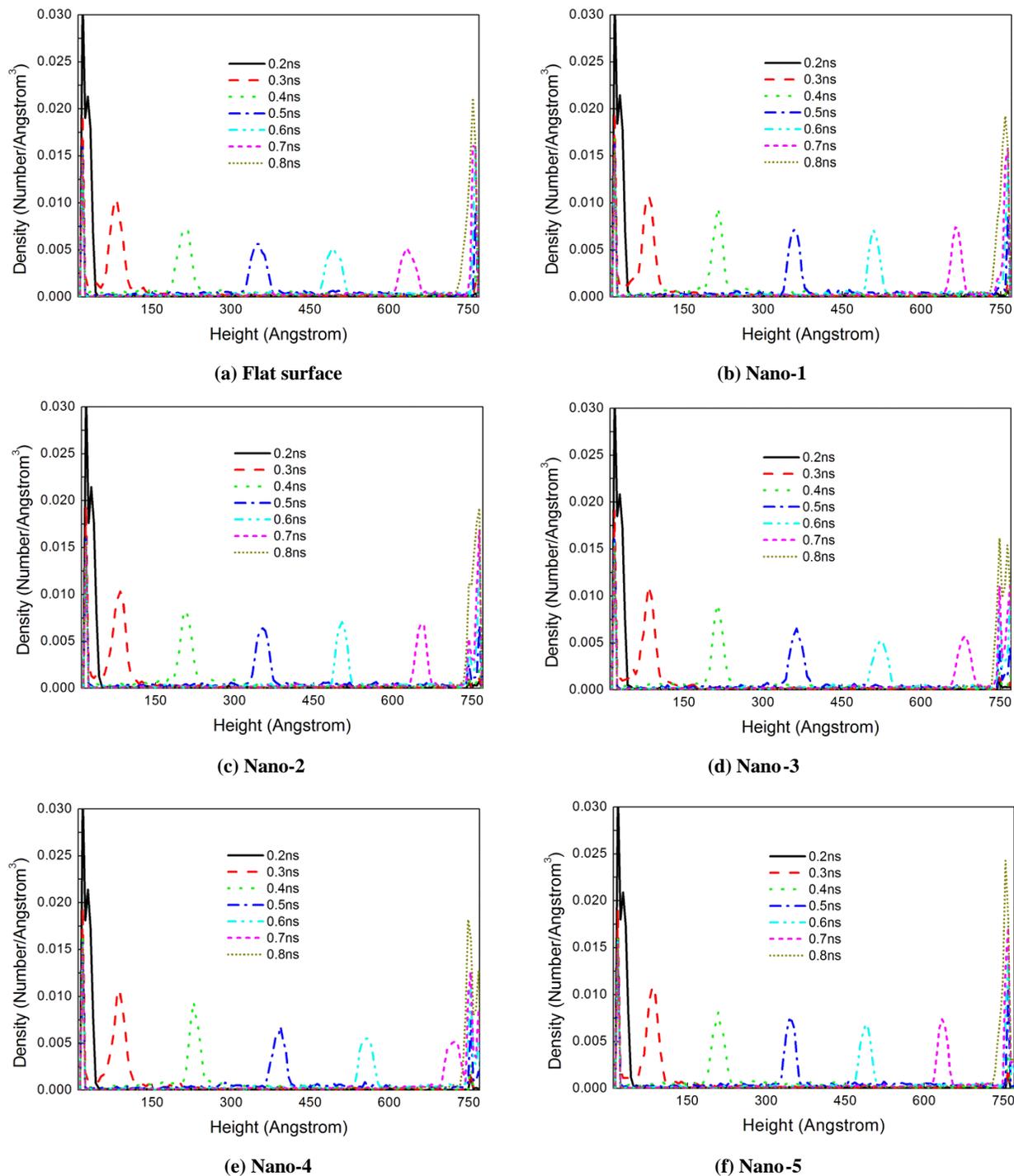

Fig. 13. Number densities of argon atoms for Nano-i structures for hot wall ($300\,K$) versus cold wall ($85\,K$). (a) Flat surface, (b) Nano-1, (c) Nano-2, (d) Nano-3, (e) Nano-4, (f) Nano-5

## 4. Conclusions

Molecular dynamic simulation was used to simulate the nanoscale vaporization and condensation. Two temperatures ($130K$ and $300\,K$) of the hot wall were set to investigate the normal and explosive vaporization on the hot wall. Nine types of cases (include one flat surface and eight kinds of nanostructures) were designed on the cold wall to study the effect of nanostructures on condensation. Copper was chosen as the material of hot wall and cold wall. Argon at liquid and vapor states were filled between the hot and cold walls as the media of heat transfer. Comparing with the normal vaporization, the trajectories of argon atoms and temperatures of argon demonstrate that the separation between liquid argon and hot wall in explosive vaporization seems not good for the transportation of thermal energy from the hot side to the cold side. Moreover, the simulation results indicate nanostructure helps to enhance condensation, because of the increase of contact surface area. The surface area including nanostructure is not the larger the better. The efficiency of heat transfer is still affected by the location of nanostructure, especially the space between each surface. There is an optimal scheme for the space between surfaces including nanostructure and the flat surface.

# Acknowledgment

The financial support for this research project from the 111 Project No. B12034 and US National Science Foundation under grant number CBET-1404482 is gratefully acknowledged.

# References

[1] D. G. Cahill, W. K. Ford, K. E. Goodson, G. D. Mahan, A. Majumdar, H. J. Maris, R. Merlin S. R. Phillpot, Nanoscale thermal transport. J. Appl. Phys. 93 (2003) 793-818.

[2] D. G. Cahill, P. V. Braun, G. Chen, D. R. Clarke, S. Fan, K. E. Goodson, P. Keblinski, W. P. King, G. D. Mahan, A. Majumdar, H. J. Maris, S. R. Phillpot, E. Pop, L. Shi, Nanoscale thermal transport. II. 2003–2012. Appl. Phys. Rev.1 (2014) 011305.

[3] J. Tigner, M. M. Sedeh, T. Sharpe, A. Bufford T. Floyd-Smith, Analysis of a platform for thermal management studies of microelectronics cooling methods. Appl. Therm. Eng. 60 (2013) 88-95.

[4] R. Hopkins, A. Faghri, D. Khrustalev, Flat Miniature Heat Pipes With Micro Capillary Grooves. J. Heat Transf. 121 (1999) 102–109.

[5] Y. J. Chen, P. Y. Wang, Z. H. Liu, Y. Y. Li, Heat transfer characteristics of a new type of copper wire-bonded flat heat pipe using nanofluids. Int. J. Heat Mass Transf. 67 (2013) 548–559.

[6] F. Lefèvre, J.-B. Conrardy, M. Raynaud, J. Bonjour, Experimental investigations of flat plate heat pipes with screen meshes or grooves covered with screen meshes as capillary structure. Appl. Therm. Eng. 37 (2012) 95-102.

[7] X. Lu, T.-C. Hua, Y. Wang, Thermal analysis of high power LED package with heat pipe heat sink. Microelectronics J. 42 (2011) 1257-1262.

[8] C. Ding, G. Soni, P. Bozorgi, B. D. Piorek, C. D. Meinhart, N. C. MacDonald, A Flat Heat Pipe Architecture Based on Nanostructured Titania. J. Microelectromechanical Syst. 19 (2010) 878-884.

[9] B. J. Alder, T. E. Wainwright, Studies in Molecular Dynamics. I. General Method. J. Chem. Phys. 31 (1959) 459-466.

[10] J. Zhang, F. Leroy, F. Müller-Plathe, Evaporation of nanodroplets on heated substrates: a molecular dynamics simulation study. Langmuir. 29 (2013) 9770-9782.

[11] A. K. M. M. Morshed, T. C. Paul, J. A. Khan, Effect of nanostructures on evaporation and explosive boiling of thin liquid films: a molecular dynamics study. Appl. Phys. A. 105 (2011) 445-451.

[12] H. R. Seyf, Y. Zhang, Molecular dynamics simulation of normal and explosive boiling on nanostructured surface. J. Heat Transf. 135 (2013) 121503.

[13] H. R. Seyf, Y. Zhang, Effect of nanotextured array of conical features on explosive boiling over a flat substrate: A nonequilibrium molecular dynamics study. Int. J. Heat Mass Transf. 66 (2013) 613-624.

[14] Y. Mao, Y. Zhang, Molecular dynamics simulation on rapid boiling of water on a hot copper plate. Appl. Therm. Eng. 62 (2014) 607-612.

[15] H. Hu, Y. Sun, Effect of nanopatterns on Kapitza resistance at a water-gold interface during boiling: A molecular dynamics study. J. Appl. Phys. 112 (2012) 053508.

[16] P. Yi, D. Poulikakos, J. Walther, G. Yadigaroglu, Molecular dynamics simulation of vaporization of an ultra-thin liquid argon layer on a surface. Int. J. Heat Mass Transf. 45 (2002) 2087-2100.

[17] J. Yu, H. Wang, A molecular dynamics investigation on evaporation of thin liquid films. Int. J. Heat Mass Transf. 55 (2012) 1218-1225.

[18] S. C. Maroo, J. N. Chung, Heat transfer characteristics and pressure variation in a nanoscale evaporating meniscus. Int. J. Heat Mass Transf. 53 (2010) 3335-3345.

[19] C. S. Wang, J. S. Chen, J. Shiomi, S. Maruyama, A study on the thermal resistance over solid-liquid-vapor interfaces in a finite-space by a molecular dynamics method. Int. J. Therm. Sci. 46 (2007) 1203-1210.

[20] G. Nagayama, M. Kawagoe, A. Tokunaga, T. Tsuruta, On the evaporation rate of ultra-thin liquid film at the nanostructured surface: A molecular dynamics study. Int. J. Therm. Sci. 49 (2010) 59-66.

[21] S. Plimpton, Fast parallel algorithms for short-range molecular dynamics. J. Comput. Phys. 117 (1995) 1-42.

[22] W. Humphrey, A. Dalke, K. Schulten, VMD: Visual molecular dynamics. J. Mol. Graph. 14 (1996) 33–38.

[23] H. J. C. Berendsen, J. P. M. Postma, W. F. van Gunsteren, A. DiNola, J. R. Haak, Molecular dynamics with coupling to an external bath. J. Chem. Phys. 81(8) (1984) 3684–3690.

[24] M. S. Zabaloy, V. R. Vasquez, E. A. Macedo, Description of self-diffusion coefficients of gases, liquids and fluids at high pressure based on molecular simulation data. Fluid Phase Equilib. 242 (2006) 43–56.

[25] H. Watanabe, N. Ito, C. K. Hu, Phase diagram and universality of the Lennard-Jones gas-liquid system. J. Chem. Phys. 136 (2012) 204102.

[26] Jia, T., Zhang, Y., Ma, H. B., Chen, J. K., Investigation of the characteristics of heat current in a nanofluid based on molecular dynamics simulation. Appl. Phys. A. 108 (2012) 537–544.

[27] K. Nakanishi, K. Toukubo, Molecular dynamics studies of Lennard-Jones liquid mixtures. V. Local composition in several kinds of equimolar mixtures with different combining rule. J. Chem. Phys. 70 (1979) 5848-5850.

[28] P. H. Hünenberger, Thermostat Algorithms for Molecular Dynamics Simulations. Adv. Polym. Sci. 173 (2005) 105–149.

[29] S. Nosé, Constant-temperature molecular dynamics. J. Phys. Condens. Matter. 115 (1990) SA115–SA119.

[30] S. Nosé, A molecular dynamics method for simulations in the canonical ensemble. Mol. Phys.100 (2002) 191-198.

[31] C. Y. Ji, Y. Y. Yan, A molecular dynamics simulation of liquid-vapour-solid system near triple-phase contact line of flow boiling in a microchannel. Appl. Therm. Eng. 28 (2008) 195-202.